\documentclass[11pt,twoside]{article}
\usepackage{asp2010}
\newcommand{\ind}[1]{_{\mathrm{#1}}}

\resetcounters

\begin{document}

\title{Testing Asteroseismology with red giants in eclipsing binary and multiple-star systems}
\author{Patrick~Gaulme$^1$
\affil{$^1$Department of Astronomy, New Mexico State University, P.O. Box 30001, MSC 4500, Las Cruces, NM 88003-8001, USA}}

\begin{abstract}
Red-giant stars are proving to be an incredible source of information for testing models of stellar evolution, as asteroseismology has opened up a window into their interiors. Such insights are a direct result of the unprecedented data from space missions CoRoT and \textit{Kepler} as well as recent theoretical advances. Eclipsing binaries are also fundamental astrophysical objects, and when coupled with asteroseismology, they would provide two independent methods to obtain masses and radii and exciting opportunities to develop highly constrained stellar models.  \citet{Gaulme_2013} reported the discovery of 13 \textit{bona fide} candidates (12 previously unknown) to be eclipsing binaries, one to be an non-eclipsing binary with tidally induced oscillations, and 10 more to be hierarchical triple systems, all of which include a pulsating red giant. When ground-based support in terms of atmospheric abundance and radial velocities are completed,  these red giants in eclipsing binary systems have the potential to become some of the most accurately studied stars.

\end{abstract}

\section{Testing asteroseismic scaling laws}
The objective of asteroseismology is to improve our knowledge about stellar properties to better understand how stars are structured and evolve. More specifically, its aim is to test whether theoretical models adequately reproduce real stars and if not, how the physics constituting the current models should be improved. Inaccurate modeling should be reflected in a failure to match the observed parameters, such as mode frequencies. The simplest analysis of asteroseismic data is based on the overall properties of the oscillations, which are the maximum amplitude frequency of the modes $\nu\ind{max}$, as well as the mean frequency separation $\Delta\nu$ between successive modes of a given degree. $\Delta\nu$ is related to the sound travel time across the stellar diameter (acoustic radius), and is therefore proportional to the square root of the mean density $\bar\rho$ to a good approximation \citep{Ulrich_1986}. $\nu\ind{max}$ is related to the surface gravity and effective temperature, and thus almost corresponds to the acoustic cut-off frequency \citep{Brown_1991,Belkacem_2011}.
Based on these assumptions, \citet{Kjeldsen_Bedding_1995} expressed the stellar radius $R$ and mass $M$ in a scaling relation with respect to the solar parameters, as function of $\Delta\nu$, $\nu\ind{max}$ and effective temperature  $T_{\rm{eff}}$:
\begin{eqnarray}
\frac{R}{R_\odot} & = & \left(\frac{\nu_{\rm{max}}}{\nu_{\rm{max},\odot}}\right) \left(\frac{\Delta\nu}{\Delta\nu_\odot}\right)^{-2} \left(\frac{T_{\rm{eff}}}{T_{\rm{eff},\odot}}\right)^{1/2}\\
\frac{M}{M_\odot} & = & \left(\frac{\nu_{\rm{max}}}{\nu_{\rm{max},\odot}}\right)^3 \left(\frac{\Delta\nu}{\Delta\nu_\odot}\right)^{-4} \left(\frac{T_{\rm{eff}}}{T_{\rm{eff},\odot}}\right)^{3/2}
\end{eqnarray}
Thus, provided that $T_{\rm{eff}}$ is known, both radius and mass can easily be retrieved from the oscillation spectrum, without having to perform a detailed mode frequency analysis. This technique has been applied to huge samples of main-sequence and red-giant stars, or so-called \textit{ensemble} asteroseismology (e.g. \citealt{Chaplin_2011c}, \citealt{Huber_2011}).
However, since both $\nu\ind{max}$ and $\Delta\nu$ are present at quite high powers, particularly for the mass, the results are sensitive to uncertainties in these observed quantities. Furthermore, the reliability of the mass and radius estimates from these relations depends on the validity of the scaling laws themselves. 

Given the importance of these scaling laws, many recent efforts have been carried out to test their validity. We may distinguish two kinds of approaches: those based on validating the relation between $\Delta\nu$ and $R$ from models and simulated data, the others based on measuring $R$ independently from asteroseismology. \citet{Stello_2009b} present the result of a wide ``hare-and-hounds'' campaign that was organized by the AsteroFLAG working group to prepare for the interpretation of \textit{Kepler} data. It consisted of two groups, the first simulating \textit{Kepler} data for a set of stars created with stellar evolution codes, the second group analyzing the simulated data to retrieve $\Delta\nu$ and $R$ from the scaling. They obtained an agreement to about 3\,\% on the radii. Such  work is important to ensure that there is no processing bias in estimating $\Delta\nu$ from the data, but it does not demonstrate that the stellar model, i.e. the physics used therein, is correct. In the same way, \citet{White_2011} focused on testing how accurately the scaling between $\Delta\nu$ and the mean stellar density $\bar\rho$ is followed by theoretical considerations. They showed that $\Delta\nu$ is indeed proportional to $\sqrt{\bar\rho}$ up to a precision of 5\,\% for stars of temperature between 4700 and 6700 K, but that a temperature-dependent term has to be taken into account for a refined estimate of $\bar\rho$, thus $R$. The same conclusions were obtained by \citet{Huber_2011} from the analysis of 1700 oscillation spectra of stars ranging from main-sequence to red-giant phases. They provided  evidence that the well-known power law connecting $\Delta\nu$ to $\nu\ind{max}$ has to be corrected for second-order effects related to $T\ind{eff}$. In addition, they find  fair agreement when comparing the measured $\Delta\nu$ and $\nu\ind{max}$ with those obtained from model evolutionary tracks, suggesting that the radii estimates from scaling are reliable to a few percent. More recently, \citet{Huber_2012} presented the measurement of five stellar radii with the CHARA interferometric array, for which asteroseismic parameters are known. The agreement is better than 4\,\%. Finally, \citet{Silva_Aguirre_2012} applied the infrared flux method \citep{Blackwell_Shallis_1977} to estimate stellar apparent diameters of 22 main-sequence stars, whose physical properties were obtained by asteroseismic measurements, \textit{Hipparcos} parallaxes, and spectrometric measurements. Their two independent distance determinations agree to better that 5\,\%, suggesting that radii estimates from asteroseismic scaling are good to about such a level.

The above-mentioned works all indicate that radius estimates from asteroseismic scaling relations are precise up to a few percent. On the contrary, similar tests of mass determination for individual stars have not been possible so far. We may note that most of the studies focused on the reliability of the $\Delta\nu$-$\bar\rho$ scaling and not on $\nu\ind{max}$, because this observable has no fully secure theoretical basis, since it is not yet possible to make reliable predictions of the amplitude of stochastically excited modes and their dependence with frequency \citep{Christensen-Dalsgaard_2012}. The dependence of the third power of the mass to $\nu\ind{max}$ makes  the need of measuring masses of stars where solar-like pulsations are detected independently from asteroseismology crucial. 

\section{Support from eclipsing binary systems }

Occasionally, it is possible to determine masses and/or radii of stars constituting an eclipsing binary or a multiple-star system, from a combination of  photometry, Doppler spectrometry, and abundance measurements. We consider only systems where the angular separation between their components is not optically resolved, since none of the targets we next present have been resolved so far. If stellar spectral lines are detected to track the Doppler shifts along their orbits, such systems are called \textit{spectroscopic binaries}. This category is divided into double-lined spectroscopic binaries (SB2), where spectral lines from both stars are visible, and single-lined spectroscopic binaries (SB1), where the spectrum of only one of the stars is seen. 

The SB2 represents the ideal case for determining the physical parameters of a system. Detailed light curve modeling allows us to determine the stellar radii relative to the orbital semi-major axis, the orbital period, the eccentricity, the orbital plane's inclination, the argument of the periastron, the ratio of effective temperatures, and the stellar albedos. Radial velocity measurements bring an additional constraint on the orbital eccentricity and argument of the periastron, and above all, lead to determine both masses and the semi-major axis. Once the orbital velocity of each star is determined, the eclipse duration gives access to the radii. Finally, abundance estimates from spectrometry give access to effective temperatures and metallicity. Nevertheless, SB1 systems where ellipsoidal and/or beaming effects are detectable constitute the second best case. If spectral lines are detected for only one star, only the sum of the masses can be deduced from radial velocities. However, in close-in binary systems, it is possible to extract the masses provided that  the tidal forces they apply on each other are enough to distort the stellar shapes or that their orbital velocities are large enough to generate relativistic beaming.

\begin{figure}
\includegraphics[width=1\textwidth]{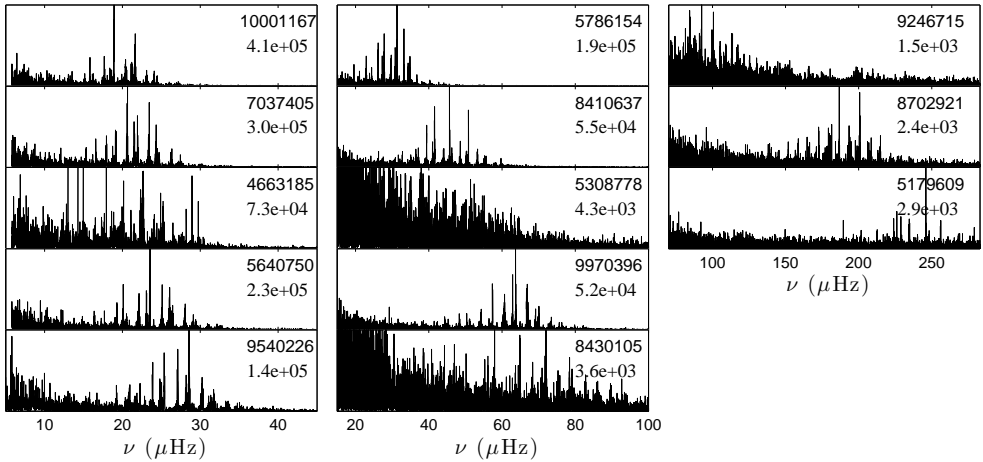}
  \caption{Power density spectra (PDS) of the 13 light curves in which global, solar-like oscillations are detected, sorted by increasing $\nu\ind{max}$ from top to bottom and left to right. Note the different frequency $x$-axis scales in each column. Each system is labeled by its \textit{Kepler} ID number. Below each KIC number, the y-axis indicates the y-range in ppm$^2$~$\mu$Hz$^{-1}$.\label{fig_1}}
\end{figure}

\begin{figure}
\includegraphics[width=1\textwidth]{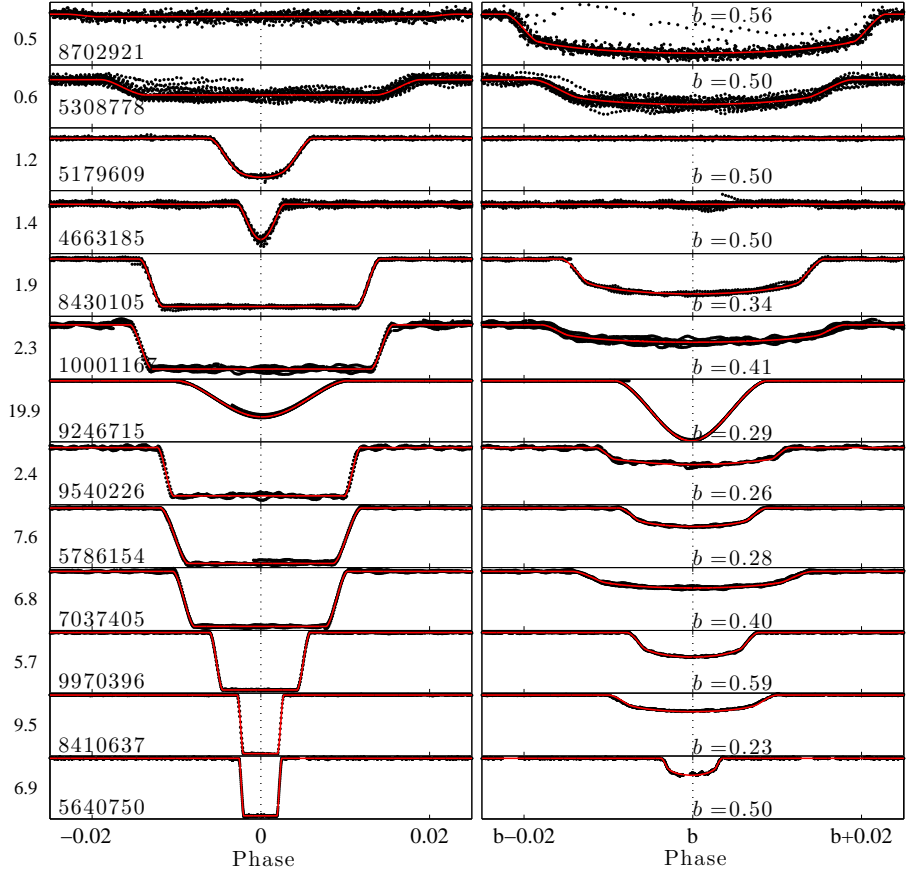}
\caption{Modeled light curves for the 13 \textit{bona fide} eclipsing binaries with a pulsating red giant \citep{Gaulme_2013}. In each case, the secondary eclipse (defined here as the red giant eclipsing the secondary star) has been set to an orbital phase of zero. The primary eclipses (secondary star eclipsing the red giant) have also been aligned, and the given $b$ value indicates the phase of the primary eclipse with respect to the secondary. (An eclipsing binary with a circular orbit would have $b = 0.5$.) The y-axis for each panel indicates the range of the normalized relative flux in percent.}
\label{fig_2}
\end{figure}

\begin{figure}
\includegraphics[width=1\textwidth]{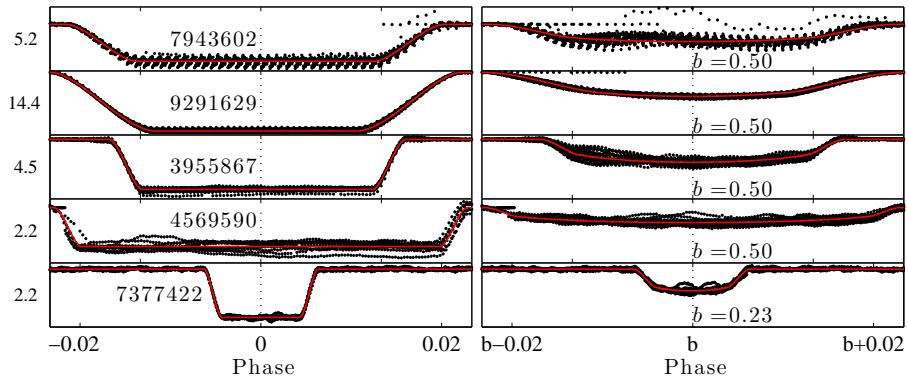}
\caption{Same as Fig. \ref{fig_2} for the five longest-period RG/EBs with no RG pulsations. }
\label{fig_3}
\end{figure}

With the \textit{Kepler} satellite, since most stars are observed at about 29-min cadence, global modes of main-sequence solar-like stars are not accessible; however, global modes of red-giant stars larger than $3.5~R_\odot$ are accessible \citep{Mosser_2012b}. \citet{Gaulme_2013} established a list of red giant candidates that likely belong to eclipsing binaries or multiple-star systems, obtained from the cross-correlation of both eclipsing binaries and red giant public catalogs. They tested whether these candidates are part of eclipsing binaries or multiple systems and characterized their main physical properties, to conclude that 13 (of 70) systems are true pulsating red giants in eclipsing binaries (hereafter RG/EBs, Fig. \ref{fig_1} \& \ref{fig_2}), while an additional five are likely RG/EBs even though no pulsations are detected so far (Fig. \ref{fig_3}). In addition, it is likely that at least 11 other systems belong to three-body configurations  composed of a pair of eclipsing main-sequence stars and a red giant. The systems span a range of orbital eccentricities, periods, and spectral types F, G, and M for the red-giant companion. One case even suggests an eclipsing binary composed of two red-giant stars and another of a red giant with a $\delta$-Scuti star. For the 7 out of the 13 RG/EBs that display the shortest orbital periods ($\leq 120$ days), phases effects are detected. For systems with larger orbital periods, the number of orbits in the current  \textit{Kepler} data set (Q0-Q13) is not sufficient to observe phase effects in the folded data.

Spectroscopic measurements from the ground will certainly help in understanding these systems too, by providing a third method for mass determination. During 2012, we started observations of the 13 RG/EBs with the ARCES \'echelle spectrometer (resolution $R = 30\,000$) at the Apache Point Observatory (APO), New Mexico. Some targets from this sample have also been observed by the APOGEE spectrometer on the SLOAN Digital Sky Survey telescope at APO within the APOKASC program to support \textit{Kepler} observations in asteroseismology, coordinated by teams from the APOGEE project and the Kepler Asteroseismic Science Consortium (KASC). 

\section{Prospects}
Hitherto, the measurement of both stellar masses and radii has not been possible for any of the stars displaying solar-like oscillations, preventing us from calibrating asteroseismology. The next objective is to focus on the 13 red giants in eclipsing binaries that have been detected so far. The second step will consist of systematically search for oscillation spectra among the eclipsing binary candidates which do not belong to the RG list. Indeed, some cool red giants or subgiant stars with $\nu\ind{max}\leq \nu\ind{Nyquist}$ certainly belong to eclipsing binaries. The last step consists of doing the same with non-eclipsing binary systems that were identified through their ellipsoidal effects only \citep{Prsa_2011}, for which one case has already been identified by \citet{Gaulme_2013}. We expect the total sample of stellar binary systems with a pulsating star that we will analyze from the \textit{Kepler} data to be between 50 and 100. Validating the mass scaling law on giant stars will also give confidence for its application to main-sequence solar-like stars that host exoplanets.

\acknowledgements The author gratefully acknowledges support from the Los Ala-mos National Laboratory Institute of Geophysics and Planetary Physics subcontract 150623-1. Part of this work was also funded by the \textit{Kepler Guest Observer Program} Cycle 2 project GO 20011, and by a NASA EPSCoR award to NMSU under contract \#~NNX09AP76A. 

\bibliography{bibi}

\end{document}